# An RXTE Search for the Sterile Neutrino Decay in Galaxy Clusters

Mark Jeffrey Henriksen

Physics Department, University of Maryland, Baltimore County, MD 21250, USA; henrikse@umbc.edu

**Abstract**

We have used long observations of galaxy clusters obtained with the Rossi X-ray Timing Explorer to search for the 3.55 keV line from sterile neutrino decay. If a lepton-number asymmetry exists in one or more types of active neutrinos in the early Universe, sterile neutrinos can be produced via the Shi–Fuller mechanism. The data consist of 11 clusters observed for a total of 3.1 megaseconds using the Proportional Counter Array. A $2.5\sigma$ excess of emission over a thermal model is found over the energy span of the 3.55 keV line in the combined spectra of the eight clusters that individually have an excess. These residuals are added to increase the signal to noise ratio of the excess, which is then modeled with a Gaussian to simulate the instrumental spectral response. We find a significant correlation (r = 0.76) for a line centered at 3.6 keV with a model flux of $3.07 \times 10^{-5}$ ph cm$^{-2}$ s$^{-1}$. Mixing angle for detected clusters ranges from 0.35 to $6.2 \times 10^{-10}$. The decay rate inferred from the line flux is strongly correlated (r = 0.87) with cluster temperature, which is due to hotter, more massive clusters having a larger amount of dark matter. Approximately half of the total flux comes from the Coma cluster. The mixing angle for Coma is calculated to be $6.2 \times 10^{-10}$. We fit the Coma cluster spectrum with two different three-component models. The first includes a Gaussian fixed at 3.55 keV to model soft emission. The flux of the Gaussian is $5.6 \times 10^{-12}$ ph cm$^{-2}$ s$^{-1}$ or 1.3% of the total flux. The second three-component model uses a second thermal component to model soft emission. This model gives a temperature of 0–17 keV for the second thermal component and a lower temperature for the hot component. This indicates that the second thermal component is modeling high-energy residuals rather than low ones, where the Gaussian is. Though our line fluxes exceed most reported detections and upper limits, they do not overproduce the dark matter. We conclude that some fraction of the marginally detected excess could be attributed to the decay line since low-temperature thermal emission and systematics fail to model it completely.

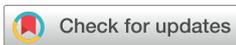

## 1. Introduction

Dark matter and dark energy are arguably the two greatest mysteries of physics beyond the standard model today. There is ample evidence from nearly a century of data utilizing multiple observationally independent methods that dark matter exists. These observations show that dark matter makes its presence known dynamically on the scales of galaxies, groups, and clusters of galaxies as well as in the overall appearance of the large-scale, filamentary universe. Direct detection of dark matter particles is elusive both astronomically and in particle accelerators as well as other types of experiments. For this reason, with few real clues as to its true identity, there are many suggestions for possible dark matter particles that can be searched for with astronomical observations. If a lepton-number asymmetry exists in one or more types of active neutrinos in the early Universe [1] this can provide flexibility in sterile neutrinos (dark matter) produced via the Shi–Fuller mechanism [2], which better matches the X-ray and structure formation constraints. Sterile neutrinos may be produced from active neutrinos via the MSW resonant conversion process. This requires a lepton asymmetry near the era of big bang nucleosynthesis. Only neutrinos that evolve through resonances adiabatically are efficiently converted. The resonant process can produce an energy distribution for sterile neutrinos that is cold in that their relatively large rest mass implies they become nonrelativistic at an early epoch, making them a cold dark matter (CDM) candidate. Figure 1 is a Feynmann diagram that shows the production of the X-ray signature. The incoming sterile neutrino couples with an active neutrino due to the small mixing angle. In the weak interaction loop, from the active neutrino a W-bozon is emitted which couples to a charged lepton. The positively charged lepton propagates and emits the X-ray. The lepton then reconnects with the W-bozon emitting the active neutrino.

Relatively recent investigations of the sterile neutrino as a candidate for dark matter have gained considerable attention due to the decay being easily accessible with astronomical X-ray detectors. The authors of Ref. [3] were the first to point out the connection between the allowable parameter space for sterile neutrino dark matter decay and the detectability in galaxies and galaxy clusters by high-resolution X-ray observatories such as Chandra and XMM. As a result there have been searches of archival and new data including Chandra [4], XMM [5], Suzaku [6], and HaloSat [7]. A comprehensive summary of these results indicates that while there are detections of a feature at the sterile neutrino decay energy, the general consensus is that its attribution to dark matter is inconclusive at best. It is pointed out that the observations are made with similar CCD detectors that may contribute to the contradictory results through variations in calibration [7].

In this paper, we investigate the spectra of 11 long observations of galaxy clusters taken with the proportional counter array (PCA) onboard the Rossi X-ray Timing Explorer (RXTE). The PCA consists of large area of gas-filled proportional counters that provide a large effective area and modest energy resolution.

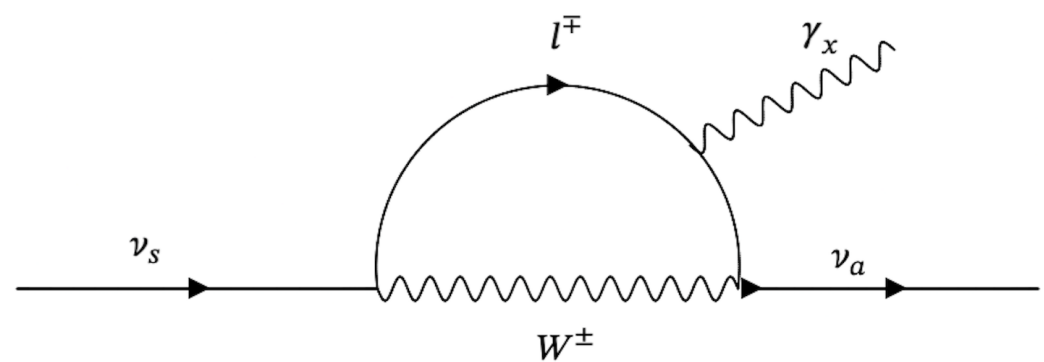

**Figure 1.** A Feynman diagram for the production of the X-ray via sterile neutrino oscillation.

## 2. Observations and Methodology

The sample of 11 clusters used in this study were all observed with the Rossi X-ray Timing Explorer (RXTE). Table 1 has basic data for these clusters including the RXTE-specific observation ID, observation length, and background subtracted count rate. Table 1 also has parameters used in model fitting: redshift and column density. The total cluster observation time is 3.1 megaseconds. The average number of source counts per cluster spectrum is 3.75 million. There are typically 57 channels in the spectrum and roughly 66,000 source counts per channel or less than 0.3% counting error. The systematic error is 0.5% [8], which dominates the counting error.

Table 1. Basic Data for the Cluster Sample.

| Cluster | ObsID | z | $n_H \times 10^{22}$ cm$^{-2}$ | Time $\times 10^5$ s | Count Rate Counts s$^{-1}$ |
|---|---|---|---|---|---|
| Abell 262 | P91149 | 0.0163 | 0.0685 | 2.11 | 2.4 |
| Abell 576 | P50199 | 0.0389 | 0.0533 | 1.44 | 2.07 |
| Abell 2256 | P60154 | 0.0581 | 0.0422 | 3.24 | 6.94 |
| Abell 496 | P40191 | 0.0329 | 0.0430 | 2.20 | 6.64 |
| Abell 1750 | P70166 | 0.0852 | 0.0269 | 1.75 | 0.77 |
| Abell 1656 | P10368 | 0.0231 | 0.0092 | 3.13 | 38.2 |
|  | P50197 |  |  | 3.80 | 36.7 |
| Abell 2319 | P40190 | 0.0557 | 0.0837 | 2.059 | 17.6 |
| Abell 3667 | P60156 | 0.0556 | 0.0426 | 1.463 | 6.40 |
| Abell 1060 | P40189 | 0.0126 | 0.0507 | 1.722 | 5.00 |
| Abell 754 | P30272 | 0.0542 | 0.0496 | 3.119 | 10.73 |
|  | P20355 |  |  | 1.206 | 10.91 |
| RXJ0658.4-5557 | P70165 | 0.296 | 0.0437 | 3.734 | 1.59 |

The following sequence of HEASoft tools are run to obtain the analysis files: pcaprepobsid, pcamergeobsids, maketime, pcaextspect2. These programs essentially run the pipeline processing in standard 2 mode with default parameters. The general sequence is to process each short observation within an obsid. Then the multiple observations are merged into a single file. Good time intervals are selected based on, for example, pointing within 6 arc min of the target, insuring at least one PCA detector is on, and the elevation is at least 4 degrees above the Earth's horizon. Source and background are extracted and deadtime correction applied. The background file utilized reflects the latest PCA background model. The response matrix is created using the appropriate gain at the time of the observation for the particular PCUs that were on. The dominant source of error is the Cosmic X-ray Background (CXB) fluctuations, which are modeled during spectral analysis.

The continuum must be accurately modeled to accurately model the decay line in the X-ray spectrum. We modeled each cluster spectrum independently to obtain the residuals around the decay line for each cluster. Each cluster must be modeled independently because each cluster is characterized by a different emission-weighted temperature, abundance, emission integral, redshift, and column density of absorbing material in the Galaxy along the cluster's observational line of sight. The broadband (2–30 keV) spectrum is modeled using a thermal (Apec) model and a power law with a high energy cutoff (cutoffpl); both models are part of the XSPEC suite of models in the HEASoft tools. The power law is used to model the fluctuations in the CXB that can occur within the PCA field of view, FWHM of $1°$. The parameters for cutoffpl are power law index 1.29, cutoff at 41.13 keV, and variable amplitude equal to 8% of the mean CXB flux at 20 keV (requires a normalization of $\pm 1.84 \times 10^{-4}$) [9]. The column density is fixed at the weighted average obtained with the HEASARC $n_H$ tool, which uses data from [10]. The redshift, CXB spectral index, and high-energy cutoff are also fixed parameters. For the thermal component, the free parameters are plasma temperature, chemical abundance [11], and emission integral. Each cluster spectrum is fit with the model described above using Xspec to obtain a minimum $\chi^2$. For the two clusters with multiple ObsIDs, Abell 754 and Abell 1656, a joint fit was performed of the two ObsIDs with an additional free parameter for the thermal normalization.

To look for emission from the 3.55 keV line from sterile neutrino decay, the residuals, obtained after accurately modeling the thermal emission for each cluster individually, were modeled to search for the signature of the decay. The PCA has an energy resolution of 18% at 6.7 keV, which is 1.2 keV. The energy calibration from the onboard Am$^{241}$ (an isotope of Americium 243) indicates that the line emission can be well-fit with a Gaussian [12]. For a line centered at 3.55 keV, the Gaussian line profile will be spread over 2.35–4.75 keV.

In standard 2 analysis, this is a channel range of 2–8 for epoch 3 observations (Abell 754, Abell 496, and Abell 1656) and 1–7 for epoch 4 and 5, which applies to the other 8 clusters in the sample. A maximum systematic error in the energy scale calibration of 0.5–2% [12] is not accounted for as it does not change the energy bins used, which extend ∼4% beyond the energy range of interest.

The delchi residuals, (data-model)/error, in each of the 7 PHA channels from each cluster are added together. The three clusters with channels 2–8 are matched as closely as possible in energy to 1–7 and are added to them. This creates a stacked 7-channel data set. A Gaussian function, as shown in Equation (1), is fit to these data with the independent variable being the mid-channel energy and the dependent variable the summed delchi residuals. The mid-channel energies are (in keV) 2.26, 2.67, 3.07, 3.48, 3.88, 4.29, and 4.69. Delchi is chosen so that each excess compared to the model is weighted by its statistical error. The free parameters for the Gaussian are the normalization, A, the line center, $\mu$, and the line width, $\sigma$. A 0.5% systematic error, which dominates the counting error, is used to weight each data point.

$$f(x) = A e^{-\left(\frac{x-\mu}{\sigma}\right)^2} \tag{1}$$

## 3. Results

### 3.1. A Soft Excess in the Co-Added Residuals

The 90% confidence limits for the Apec model free parameters are given in Table 2 for each cluster. Those are temperature in keV, abundance in fraction of Solar, and normalization in $cm^{-5}$. The best fit values with 95% confidence bounds for the Gaussian fit to the delchi residuals that span the 3.55 keV line are given in Table 3.

**Table 2.** Model Fitting Results: Thermal and CXB.

| Cluster | kT keV | Abundance Fraction of Solar | Thermal Normalization $cm^{-5}$ |
|---|---|---|---|
| Abell 262 | 2.05–2.24 | 0.23–0.37 | 0.066–0.076 |
| Abell 576 | 3.47–3.88 | 0.21–0.34 | 0.028–0.032 |
| Abell 2256 | 6.66–6.96 | 0.20–0.27 | 0.067–0.071 |
| Abell 496 | 4.08–4.23 | 0.28–0.33 | 0.164–0.177 |
| Abell 1750 | 3.27–3.97 | 0.13–0.37 | 0.014–0.019 |
| Abell 1656 | 7.45–7.55 | 0.21–0.23 | 0.290–0.296 |
|  |  |  | 0.300–0.303 |
| Abell 2319 | 8.84–9.12 | 0.20–0.27 | 0.141–0.147 |
| Abell 3667 | 6.59–6.90 | 0.18–0.25 | 0.081–0.086 |
| Abell 1060 | 3.33–3.51 | 0.23–0.29 | 0.084–0.091 |
| Abell 754 | 9.00–9.24 | 0.20–0.25 | 0.088–0.091 |
|  |  |  | 0.086–0.089 |
| RXJ0658.4-5557 | 10.23–11.79 | 0–0.35 | 0.019–0.022 |

**Table 3.** The 95% Confidence Limits for the Gaussian Fit Parameters.

| Parameter | Lower Bound | Best Fit | Upper Bound |
|---|---|---|---|
| A (normalization) | 0.0 | 0.7 | 1.5 |
| $\mu$ (line center in keV) | 3.1 | 3.6 | 4.1 |
| $\sigma$ (line width in keV) | 0.2 | 0.7 | 1.4 |

The R parameter, which characterizes goodness of fit, is 0.76, indicative of a strong correlation. The best fit line center for the Gaussian is 3.6 keV with a 95% error of 0.5 keV. Figure 2 shows the Gaussian fit to the delchi residuals. The model flux is $3.07 \times 10^{-5}$ ph cm$^{-2}$ s$^{-1}$. The feature is 2.5$\sigma$, below the standard detection threshold, and is presented with that caveat. This significance is obtained by adding up the deviations from the model measured in units of $\sigma$.

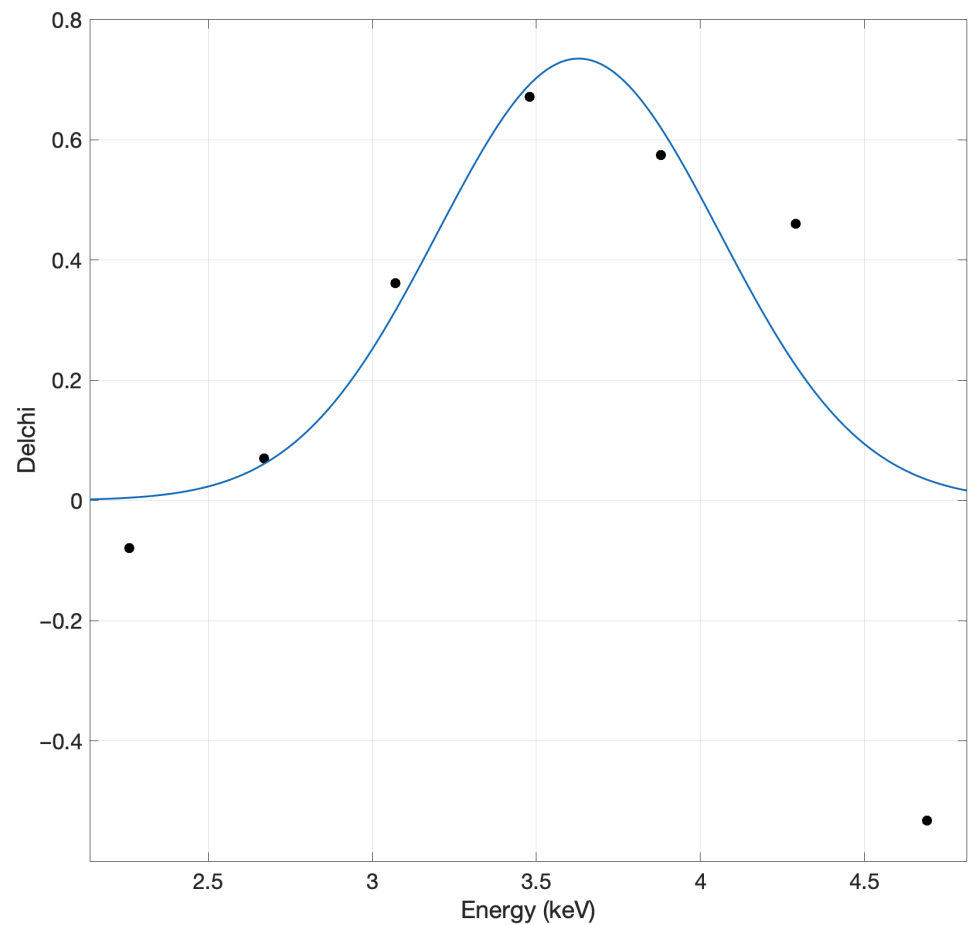

**Figure 2.** A Gaussian function is fit to a sum of the delchi or (data-model)/error in each channel covering the energy range of the 3.55 keV decay feature folded through the RXTE PCA energy resolution. A 0.5% error dominates the counting error and is used as a weight on each fit point. The error is approximately the size of the data point plotted. The R value obtained from the best fit is 0.76.

The lowest energy channel at 2.26 keV and the two highest at 4.29 and 4.69 keV are the most divergent.

Figure 3 shows the quantity (data/model) for each cluster, represented by a different color. The model is the combined thermal and CXB model. A ratio of 0.995–1.005 brackets the systematic error of the PCA of 0.5% [8]. For the lowest channels the distribution is quite broad, indicating those channels provide the most ambiguous data. The standard deviation of the ratios for the channels between clusters (0.0785, 0.0339, 0.0148, 0.0114, 0.0088, 0.0123, 0.0099) is generally much smaller in the five higher energy channels. This indicates that the higher energy channels that contribute to the Gaussian line feature are not dominated by systematics in the higher energy channels. Perhaps the large spread in the lowest energy channel is indicative of unknown calibration errors at low energy. As is apparent in Figure 2, it does not contribute to the excess.

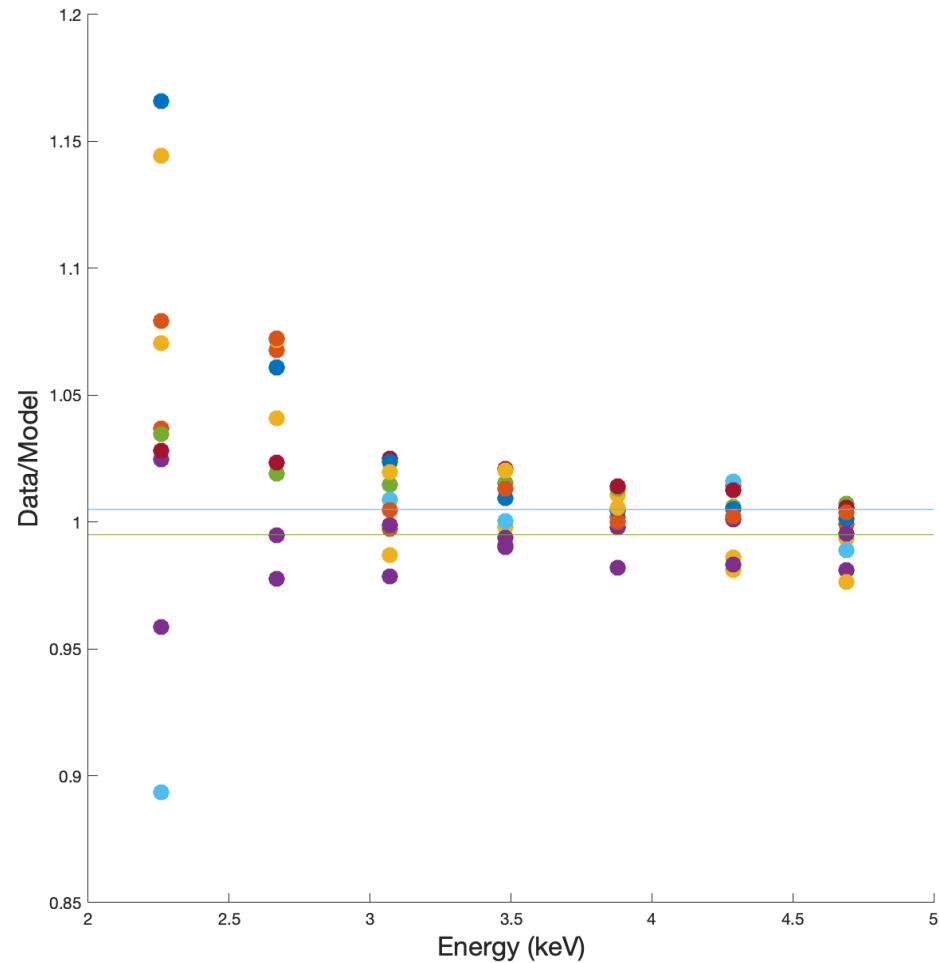

**Figure 3.** The ratio (data/model) in each channel covering the energy range of the 3.55 keV decay feature folded through the RXTE PCA energy resolution is shown for each cluster. Lines are shown that bracket the systematic error for the PCA as ±∼0.5%. The features can be compared with the systematic error and the number of clusters driving the excess evaluated as shown in the text.

*3.2. Multi-Component Spectral Modeling*

We model the broadband Coma cluster spectrum to fit the 3.55 keV line and the thermal continuum at the same time, in contrast to fitting residuals. Coma is singled out because it has a higher count rate than all of the other clusters combined. The model consists of three components: an Apec model for the thermal emission, Gaussian for sterile neutrino decay emission, and cutoffpl for the hard X-ray background fluctuations (CXB). The Coma cluster observational data is shown in Table 1 and consists of two observation periods separated by 17 months. Each Obsid has independent normalization for the thermal component as well as the CXB. All other parameters are shared by the obsids: $n_H$, kT, abundance, redshift, Gaussian line center, width, and normalization. The fixed parameters are $n_H$ (at $9.2 \times 10^{19}$ cm$^{-2}$), redshift (at 0.0231), line center (at 3.55 keV), and width (at 1.2 keV). The line width corresponds to the PCA spectral resolution at 3.55 keV. The CXB is modeled as described earlier. The results are shown in Tables 4 and 5. The thermal component normalization, which is proportional to the emission integral, for the ObsIDs is in excellent agreement, justifying the sharing of some physical parameters.

**Table 4.** The 90% Confidence Limits for the Three-Component Fit To Coma.

| kT (keV)  | Abundance | 10368 Thermal Norm | 50197 Thermal Norm | Gaussian Norm    |
|-----------|-----------|--------------------|--------------------|------------------|
| 8.18–8.38 | 0.24–0.28 | 0.28–0.29          | 0.28–0.30          | 0.00058–0.0015   |

**Table 5.** The 90% Confidence Fluxes ($\times 10^{-10}$ ergs cm$^{-2}$ s$^{-1}$) for the Three-Component Fit to Coma.

| 10368 CXB | 50197 CXB | 10368 Thermal | 50197 Thermal | Gaussian |
|-----------|-----------|---------------|---------------|----------|
| −0.28     | 0.028     | 4.20          | 4.26          | 0.056    |

*3.3. Attribution of Gaussian Component to Sterile Neutrino Decay*

The X-ray luminosity is calculated from the observed soft flux and the luminosity distance for each cluster. We assume that the residuals are due to the 3.55 keV line and calculate the soft flux from them. The residuals are the difference between the observed count rate and the model (thermal plus CXB)-predicted count rate in each channel spanning the PCA convolved line. All channels are then added to get the total count rate. The total count rate is converted to flux using a conversion from WebSim for a source with a Gaussian at 3.55 kev observed with the top layer of the PCA. The conversion is $3.368 \times 10^{-12}$ ergs cm$^{-2}$ s$^{-1}$ per PCA cnt s$^{-1}$. The conversion to photon flux is made assuming that all of the photons are emitted at 3.55 keV. The soft flux ($F_{DM}$) and soft luminosity (LDM) are given in Table 6. When the data is below the thermal model, a negative line flux results, as shown in the table. Physically this is no line emission or zero mixing angle.

**Table 6.** Calculation of Mixing Angle.

| Cluster       | x    | $M_{MD}$ $10^{14}$ Solar Masses | $D_L$ Mpc | $F_{DM}$ $10^{-6}$ ph cm$^{-2}$ s$^{-1}$ | LDM $10^{48}$ ph s$^{-1}$ | $\sin^2(2\Theta)$ $\times 10^{-10}$ |
|---------------|------|--------------------------------|-----------|-----------------------------------------|---------------------------|-------------------------------------|
| A262          | 4.8  | 1.92                           | 70        | 11.18                                   | 6.18                      | 0.35                                |
| A576          | 11.2 | 7.81                           | 175       | 16.97                                   | 58.77                     | 2.0                                 |
| A2256         | 16.3 | 20.8                           | 267       | −1.4                                    | −11.28(0)                 | -                                   |
| A496          | 9.5  | 7.47                           | 149       | 31.34                                   | 78.69                     | 2.8                                 |
| A1750         | 23.1 | 15.98                          | 407       | 1.76                                    | 32.97                     | 0.5                                 |
| Coma Cluster  | 6.76 | 9.50                           | 109       | 165.25                                  | 222.05                    | 6.2                                 |
| A1060         | 3.73 | 2.28                           | 65        | 30.05                                   | 14.18                     | 1.7                                 |
| A2319         | 15.7 | 26.73                          | 254       | 76.95                                   | 561.47                    | 5.4                                 |
| A3667         | 15.7 | 20.20                          | 255       | 14.43                                   | 106.12                    | 1.4                                 |
| A754          | 15.3 | 26.54                          | 255       | −1.12                                   | −8.23(0)                  | -                                   |
| Bullet Cluster| 63.8 | 134.3                          | 1580      | −8.80                                   | −2484.56                  | -                                   |

A linear regression is performed using the LDM and cluster temperature for the clusters with Gaussian component detections. The R value is 0.87, indicating a strong correlation. The results are given in Table 7 and shown in Figure 4. The correlation implies that hotter, more massive clusters have a higher X-ray luminosity. Since the X-ray luminosity is indicative of the sterile neutrino decay rate, the correlation is consistent with more massive clusters having a higher decay rate, which is to be expected since they have more dark matter.

**Table 7.** The 95% Confidence Limits for the Linear Fit to Luminosity versus Temperature.

| Parameter   | Lower Bound | Best Fit | Upper Bound |
|-------------|-------------|----------|-------------|
| slope       | 30.12       | 68.07    | 106.03      |
| Y Intercept | −417.80     | −208.78  | 0.25        |

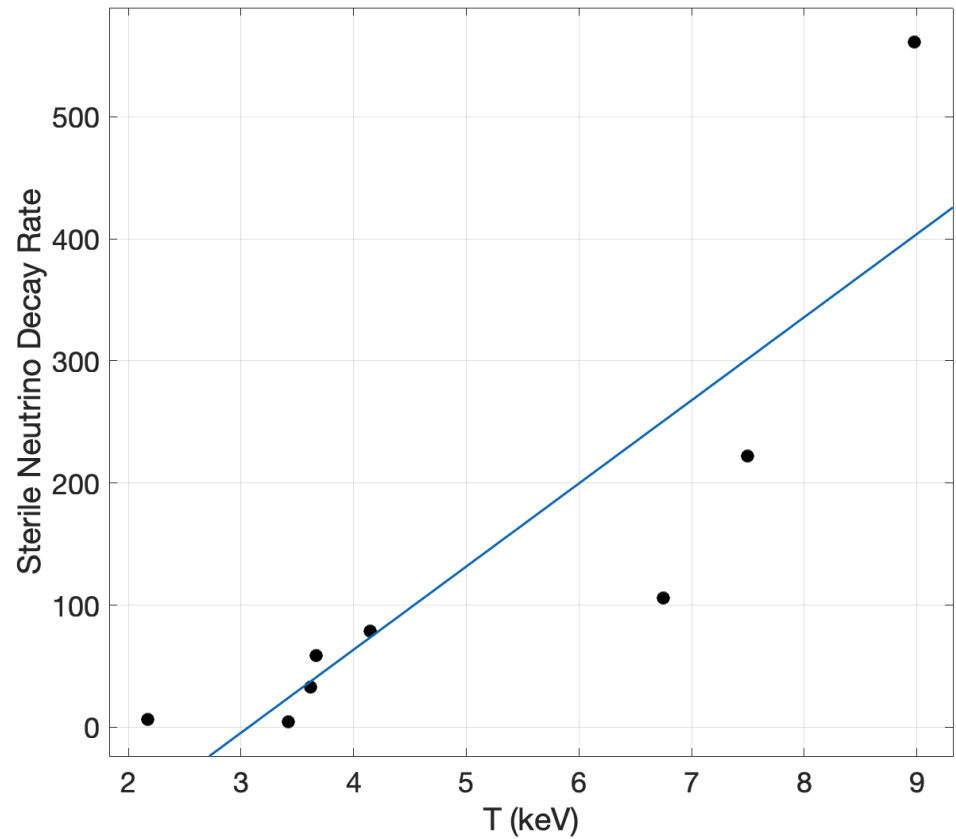

**Figure 4.** This figure shows the X-ray luminosity ($L_{DM}$ photons s$^{-1}$ given in Table 6) attributed to the 3.55 keV line from sterile neutrino decay. It is proportional to the decay rate. The correlation with temperature is strong, R = 0.87, and indicates that hotter, more massive clusters have a higher decay rate.

We looked for any possible indication that the variable CXB powerlaw might be affecting our fits. One way is to look at the number of times the CXB normalization pegs at the extreme flux allowed. This could indicate that the CXB model component is trying to fit some spectral emission not being modeled. We looked at the 11 best fit normalizations and found that only four were pegged at the extreme value. It is possible that there is non-thermal emission in some clusters at the high-energy part of the spectrum, generally above 10 keV, that the CXB component is trying to fit. We investigate this in future work. Looking at the Gaussian flux contribution of the four pegged at the extreme value compared to the unpegged shows no apparent correlation between the CXB and Gaussian contribution.

*3.4. Calculation of Mixing Angle*

The calculated mixing angle, the rate of neutrino decay, is constrained primarily by the observed X-ray photon flux and the distance to the cluster. The mixing angle also depends on cluster mass and the sterile neutrino energy. The cluster mass, $M_{DM}$, is calculated using mass equations in [13]. The dependencies are given in Equation (2).

$$M_{DM} = 1.65 \times 10^{14} M_\odot \left(\frac{0.62}{\mu}\right)\left(\frac{\delta}{1.0}\right)\left(\frac{R_c}{250\,\text{kpc}}\right)\left(\frac{T}{8.62\,\text{keV}}\right)\left(\frac{x^3}{1+x^2}\right) \quad (2)$$

We assume a canonical core radius value (250 kpc), gas density parameter $\delta = (3/2)\beta$ with canonical beta = 2/3, and a fully ionized gas with solar abundances. Gas parameters are needed because hydrostatic equilibrium is the basis of the mass equation. The important parameters in this equation are those have been found to vary the most from cluster to cluster: the number of core radii (x in Table 6) within the PCA fov (which depends on the distance to the cluster), and the temperature. The temperature is our best-fit RXTE-measured PCA cluster temperature given in Table 2. The photon flux $F_{DM}$ is for an emitted photon energy of 3.55 keV, which appears as $m_s$ in Equation (3). In general, comparison between X-ray determined masses based on hydrostatic equilibrium and masses based on gravitational lensing indicate that there is a factor ∼2 underestimation of their masses [14]. This reduces the mixing angle by approximately a factor of two, which reduces but does not remove the discrepancy with other reported values as discussed in Section 4.2. We use Equation (3) obtained from [5] for the mixing angle calculation. All of the parameters used in Equations (2) and (3) are given in Tables 1, 2 and 6 or in this section. $F_{DM}$ is the flux of the Gaussian component, $D_L$ is the distance to the cluster, Z is the redshift, and $m_s$ is the sterile neutrino particle mass, 7.1 keV, chosen to match the unidentified feature in previous X-ray studies.

$$Sin^2(2\Theta) = \frac{F_{DM}}{12.76}\Big(\frac{10^{14}M_\odot}{M_{DM}}\Big)\Big(\frac{D_L}{100_{Mpc}}\Big)^2\Big(\frac{1}{1+z}\Big)\Big(\frac{1\,\text{keV}}{m_s}\Big)^4 \qquad (3)$$

## 4. Discussion

*4.1. Possible Sources of Soft X-Ray Contamination*

Our method of co-adding residuals is preferable to the more common procedure of co-adding the data for cluster X-ray emission because there is significant variation in each cluster's thermal spectrum so that continuum and line emission vary from cluster to cluster. This variation is even more pronounced for the PCA data because of the broad PSF of the PCA. The possibility of confusing the Gaussian line emission with a low-temperature gas component is the largest source of uncertainty in our analysis and it is addressed in this section.

Though not generally constrained by the RXTE data, there are significant sources of cool gas that manifest in the X-ray spectra of clusters below 2 keV: for example, the WHIM [15], dark X-ray galaxies [16], and the brightest cluster galaxies [17]. If we adopt the upper limit on the mixing angle for 117 clusters [18], which is typical of the published upper limits, this limits the fraction of the RXTE Gaussian component attributed to DM to 7–88% for the largest Gaussian fluxes, Coma and Abell 1750, respectively. This then implies 12–93% of the emission is some other soft X-ray source for these clusters. Figure 5 shows the predicted effect of adding a low-temperature component on the X-ray spectrum. The low-temperature and Gaussian components both contribute to residuals over the hot gas cluster model. However, their spectral shape is quite different, as is apparent in Figure 5, which shows a three-component model consisting of a high-temperature thermal, a low-temperature thermal, and a Gaussian with arbitrary normalizations which allow comparison of shapes. The Gaussian clearly dominates the low-temperature component over most of its width. However, the excess emission over the high-temperature component could clearly be a mixture of the two.

A large soft component could be the WHIM or diffuse gas associated with galaxies. We model the characteristics of such a thermal component that can provide the residual emission directly. This is done by modeling the 2–30 keV PCA spectrum for Coma with a three-component model: two Apec models for hard and soft thermal emission and a cutoffpl to model the hard X-ray background. The Coma cluster is modeled because of its superior signal to noise ratio. The three-component models compared have either the soft emission modeled with a Gaussian or a thermal spectrum. This tests the hypothesis that the component is entirely soft, thermal emission. The Coma cluster observations consist of two Obsids. The normalization for each obsid is tied for each thermal component as well as the CXB. The shared model parameters are $n_H$, first kT and abundance, redshift, first kT normalization, second kT and abundance, and second kT normalization. The frozen shared parameters are $n_H$ (at $9.2 \times 10^{19}$) and redshift (at 0.0231). The other shared parameters are free. The CXB is modeled as described earlier. The results are shown in Table 8. The range of model parameters includes no soft thermal emission in contrast to the Guassian model for Coma, which is always non-zero (see Table 4). Modeling using a soft thermal component fails to adequately fit the residual soft emission and appears to model harder residuals. This is in contrast to the successful modeling with a Gaussian component.

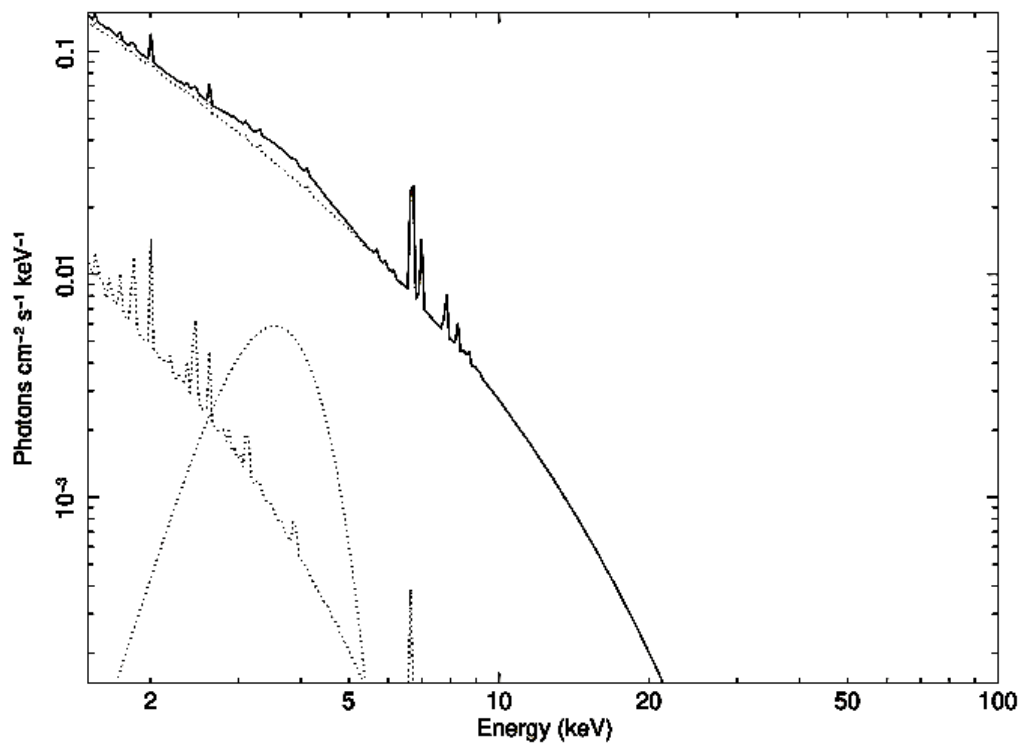

**Figure 5.** Three models are shown with nominal cluster parameters. A high-temperature Apec model with kT = 6 keV and abundance 30% Solar, a low-temperature Apec model with temperature 1.5 keV and abundance of 30% Solar, and a Gaussian with the best fit parameters: line center 3.58 keV and sigma 0.69. The normalizations are 1, 0.1, and 0.01, respectively. The Gaussian dominates the low-temperature component over most of its width. However, the excess emission over the high-temperature component could be a mixture of the two.

**Table 8.** The 90% Confidence Limits for the Two Thermal Components' Fit.

| First kT | Abundance | First kT Norm | Second kT | Second kT Norm |
|---|---|---|---|---|
| 0.0–17.2 | 0.21–0.23 | 0.010–0.29 | 7.08–7.70 | 0.27–0.30 |

### 4.2. Physicality of the RXTE Result

Active neutrinos oscillate into sterile neutrinos with a small mixing angle $\sim 10^{-7}$–$10^{-5}$. This is a non-resonant oscillation that occurs in the early Universe and produces a warm dark matter candidate. When active and sterile neutrino masses align, Shi–Fuller resonant oscillation, which requires a lepton asymmetry, produces a cold dark matter candidate, $\sim$1–10 keV, with a lower mixing angle, $\sim 10^{-10}$–$10^{-8}$. The range of mixing angles we report here is consistent with the latter range. Refs. [19,20] provide bounds on the combination of mixing angle and sterile neutrino mass that are allowed based on current theoretical and observational constraints. Extremely low values of mixing angle provide too little dark matter. Too-high values overproduce it. In addition, there are a number of X-ray upper limits that favor a mixing angle less than $10^{-11}$ for a 7.1 keV sterile neutrino. Our detections in clusters, ranging from $3.5 \times 10^{-11}$ to $6.2 \times 10^{-10}$, do not overproduce the dark matter. However, as pointed out earlier, total attribution of the excess emission to dark matter provides a mixing angle in excess of most of the X-ray upper limits obtained with other measurements.

## 5. Conclusions

We have analyzed 3.1 megaseconds of Proportional Counter Array observations of 11 clusters. We find evidence of a significant correlation (R = 0.76) for a line centered at 3.6 keV with an average flux of $3.1 \times 10^{-5}$ ph cm$^{-2}$ s$^{-1}$. The mixing angle is calculated to be $2.4 \times 10^{-9}$. The Mixing angle for the individually detected clusters ranges from 0.35 to $6.2 \times 10^{-10}$. The decay rate inferred from the line flux is strongly correlated (R = 0.87) with cluster temperature, which is due to hotter, more massive clusters having a larger amount of dark matter. Approximately half of this flux comes from the Coma cluster. We fit the Coma cluster spectrum with two different three-component models to test if the soft residuals could be attributed to thermal emission as opposed to a Gaussian line. The first includes a Gaussian fixed at 3.55 keV to model soft emission. The flux of the Gaussian is $5.6 \times 10^{-12}$ ph cm$^{-2}$ s$^{-1}$ or 1.3% of the total flux. The second three-component model uses a second thermal component to model soft emission. This model gives a temperature of 0–17 keV for the second thermal component and a lower temperature for the hot component. This indicates that the second thermal components is modeling high-energy residuals rather than low ones, where the neutrino decay would occur. Though our X-ray line fluxes exceed most reported detections and upper limits, they do not overproduce the dark matter predicted by [3,20]. We conclude that some of the Gaussian component could be thermal but not all of it. Thus some fraction of the Gaussian is attributable to the sterile neutrino decay.

**Funding:** This research received no external funding.

**Data Availability Statement:** All of the data used in this study is publicly available at the NASA HEASARC.

**Acknowledgments:** I am grateful to Keith Jahoda for making suggestions on the manuscript.

**Conflicts of Interest:** The authors declare no conflicts of interest.